\documentclass[conference,compsoc]{IEEEtran}
\IEEEoverridecommandlockouts
\usepackage{amsmath,amsfonts}
\usepackage{algorithmic}
\usepackage{algorithm}
\usepackage{array}
\usepackage[caption=false,font=normalsize,labelfont=sf,textfont=sf]{subfig}
\usepackage{textcomp}
\usepackage{stfloats}
\usepackage{url}
\usepackage{verbatim}
\usepackage{graphicx}
\usepackage{cite}
\usepackage{enumitem}
\usepackage{xcolor,colortbl} % column color in table
 \usepackage{setspace}%
\usepackage{textcomp}
\usepackage{multirow}
\newcolumntype{?}{!{\vrule width 1pt}}
\newcolumntype{P}[1]{>{\centering\arraybackslash}p{#1}}
\newcolumntype{L}[1]{>{\centering\arraybackslash}p{#1}}
\usepackage{tikz}
\usepackage{amsmath}

\newcommand{\extra}[1]{\textcolor{red}{}}
\newcommand{\toremove}[1]{\textcolor{blue}{}}
\newcommand{\mycomment}[1]{}

\def\BibTeX{{\rm B\kern-.05em{\sc i\kern-.025em b}\kern-.08em
    T\kern-.1667em\lower.7ex\hbox{E}\kern-.125emX}}
\begin{document}

\title{Environmental Rate Manipulation Attacks on Power Grid Security}

\author{\IEEEauthorblockN{Yonatan~Gizachew~Achamyeleh$^1$, Mohamad Habib Fakih$^1$, Gabriel Garcia$^1$, Anomadarshi Barua$^2$, \\ Mohammad Abdullah Al Faruque$^1$}
\IEEEauthorblockA{
{\textit{$^1$University of California, Irvine, CA, USA.}} 
{\textit{\{yachamye, mhfakih, gegarci1, alfaruqu\}@uci.edu}} \\
{\textit{$^2$George Mason University, VA, USA.}} 
{\textit{abarua8@gmu.edu}} 
}

}

\author{\IEEEauthorblockN{Yonatan Gizachew Achamyeleh, Yang Xiang, Yun-Ping Hsiao, Yasamin Moghaddas, \\ Mohammad Abdullah Al Faruque}
\IEEEauthorblockA{
{\textit{University of California, Irvine, CA, USA.}} \\
{\textit{\{yachamye, yxiang10, yunpinh, ymoghadd, alfaruqu\}@uci.edu}}
}
% \thanks{Yonatan Gizachew Achamyeleh, Yang Xiang, Yun-Ping Hsiao, Yasamin Moghaddas, and Mohammad Abdullah Al Faruque are with the University of California, Irvine, Irvine, CA 92697 USA. (e-mail: yachamye/yxiang10/yunpinh/ymoghadd/alfaruqu@uci.edu). }
}

% The paper headers
\markboth{Journal of \LaTeX\ Class Files,~Vol.~14, No.~8, August~2021}%
{Shell \MakeLowercase{\textit{Achamyeleh et al.}}: }

\IEEEpubid{}
% Remember, if you use this you must call \IEEEpubidadjcol in the second
% column for its text to clear the IEEEpubid mark.

\maketitle

\begin{abstract}
The growing complexity of global supply chains has made hardware Trojans a significant threat in sensor-based power electronics. Traditional Trojan designs depend on digital triggers or fixed threshold conditions that can be detected during standard testing. In contrast, we introduce Environmental Rate Manipulation (ERM), a novel Trojan triggering mechanism that activates by monitoring the rate of change in environmental parameters rather than their absolute values. This approach allows the Trojan to remain inactive under normal conditions and evade redundancy and sensor-fusion defenses. We implement a compact 14~$\mu$m$^2$ circuit that measures capacitor charging rates in standard sensor front-ends and disrupts inverter pulse-width modulation PWM signals when a rapid change is induced. Experiments on a commercial Texas Instruments solar inverter demonstrate that ERM can trigger catastrophic driver chip failure. Furthermore, ETAP simulations indicate that a single compromised 100~kW inverter may initiate cascading grid instabilities. 
The attack's significance extends beyond individual sensors to entire classes of environmental sensing systems common in power electronics, demonstrating fundamental challenges for hardware security.
\end{abstract}

\begin{IEEEkeywords}
Environmental Rate Manipulation, Embedded Systems, Solar Inverters, Hardware Trojans, Power Grid Security, Supply Chain Attacks
\end{IEEEkeywords}
\section{Introduction}
Power systems increasingly depend on inverter-based resources for renewable energy integration, distributed generation, and sophisticated grid services~\cite{solar_integration_doe}. However, this proliferation through complex global supply chains introduces new security vulnerabilities, directly challenging the reliability and resilience of the power grid~\cite{sharma2022complexity}. Concerns are significant; the North American Electric Reliability Corporation identified supply chain attacks as one of the most critical threats to electric utilities in their 2023 Grid Security Exercise~\cite{nerc_gridex_vii_2023}. Recent incidents, such as the compromise of electronic devices with embedded malicious circuits originating from their supply chains, have amplified these fears~\cite{guardian_2024}. Such attacks possess the dangerous capability to bypass traditional security measures. They can remain dormant until specific conditions trigger their activation, making them particularly hazardous for critical infrastructure.

Current hardware security research largely targets the detection of malicious circuits, known as hardware Trojans, by looking for specific digital sequences or extreme environmental conditions~\cite{sadeghi2015security, tehranipoor_survey_2010, skorobogatov2011physical, hu2020overview, akter2023survey}. These conventional detection methods can often be foiled through systematic testing or operational monitoring if the trigger mechanisms are known or predictable~\cite{hu2020overview}. Power system equipment, however, presents distinct security challenges. These systems have extended deployment lifetimes, limited update capabilities, and a direct interface with critical infrastructure, creating an environment where subtle attacks can thrive.

We present Environmental Rate Manipulation (ERM), a fundamentally different strategy for hardware-level attacks on power system components. Instead of using specific digital patterns or threshold violations, ERM monitors the rate of change in environmental sensor readings to initiate malicious behavior. This mechanism operates entirely within normal sensor ranges while exploiting how quickly those readings change. This approach makes ERM exceptionally difficult to detect using conventional security measures. Even systems employing redundant sensors or sensor fusion remain vulnerable because ERM does not attempt to provide false readings. It manipulates the rate of change in environmental variables affecting a single sensor, all while maintaining readings within accepted parameters.

The significance of this attack method stems from both its stealthy design and its broad applicability. The Trojan's minimal 14~$\mu$m$^2$ circuit footprint and its use of existing sensor infrastructure make it particularly difficult to identify through conventional supply chain security checks or post-manufacturing testing. While we demonstrate ERM using temperature sensors, the underlying concept can be applied to any sensor that includes signal conditioning circuitry. Examples include Hall-effect sensors for current measurement or other sensors commonly found in power electronics.

ERM is especially concerning for infrastructure that relies on environmental sensing for its operation~\cite{rao2020applications, freitas2022design, meribout2022sensor}, such as grid-tied solar inverters~\cite{al2020grid}. Using a commercial Texas Instruments solar inverter as our test platform, this paper demonstrates how ERM enables a particularly severe attack. Our hardware validation showed the attack caused rapid destruction of the photovoltaic full-bridge driver chip by manipulating PWM signals.

To address this threat comprehensively, our study conducts an end-to-end analysis of ERM's impact on solar inverters. We begin by outlining the theoretical foundations of ERM. This is followed by LTSpice~\cite{ltspice} and Simulink~\cite{simulink} simulations to demonstrate potential disruptions caused by the malicious circuit at different power conversion stages. The implications, however, extend beyond individual device failure. Through ETAP~\cite{etap} simulations of an IEEE 13 bus system~\cite{ieee13bus}, we show how a single compromised 100kW inverter can trigger cascading failures through reactive power-voltage interactions, ultimately causing grid-wide blackouts. This escalation from component-level manipulation to system-wide failure demonstrates the nature of protecting against such stealthy hardware Trojans. The main contributions of this research are:
\begin{enumerate}
    \item Development of ERM as a novel hardware Trojan triggering mechanism that differs from existing approaches by exploiting sensor rate-of-change.
    \item Implementation \& validation of a stealthy 14~$\mu$m$^2$ trigger circuit that bypasses hardware security measures.
    \item Demonstration of cascading system failures from device-level to grid-scale impacts.
    \item Analysis of broader implications for hardware Trojan attacks in sensor-based systems.
\end{enumerate}

\section{Background and Related Work}
\label{sec:background}

\subsection{Grid-Tied Solar Inverters: An Overview}
Grid-tied solar inverters transform photovoltaic energy into grid-compatible electrical power through a two-stage conversion process~\cite{al2020grid, kjaer_review_2005}. The first stage employs a DC-DC converter to stabilize and boost the solar panel output, while the second stage converts this regulated DC voltage into synchronized AC waveforms through precisely controlled Pulse Width Modulation (PWM) signals. 

These inverters incorporate MPPT algorithms to optimize energy harvesting~\cite{katche_comprehensive_2023}, while protection mechanisms rely on sensors to monitor important parameters like temperature, current, and voltage~\cite{rao2020applications, freitas2022design}.

\subsection{Security Landscape and Hardware Trojans}
Solar inverters are susceptible to various security threats that compromise their integrity and disrupt grid operations~\cite{barua_hall_2020, gursoy2021self, tuyen_comprehensive_2022, misbrener_cyberattacks_2019, vodapally2023overview}. These threats include supply chain attacks, network-based attacks that exploit communication protocols, and physical attacks that involve tampering with the inverter's hardware or firmware~\cite{tuyen_comprehensive_2022, ye_review_2022, cryar2023supply}. Researchers have also demonstrated the susceptibility of inverters to spoofing attacks, which manipulate voltage and current measurements~\cite{dan2022novel, barua_hall_2020}.

Hardware Trojans (HTs) are another insidious threat to solar inverter security.
HTs are malicious modifications or additions to electronic circuits designed to alter the normal operation of the host system~\cite{tehranipoor_survey_2010}. 
They typically consist of a trigger mechanism and a payload~\cite{tehranipoor_survey_2010}. The trigger mechanism activates the malicious payload under specific conditions while maintaining stealth. Traditional triggers fall into two main categories: digital and analog~\cite{trippel2019extensible}.

Digital triggers rely on specific patterns or conditions in the digital domain. These include time-based triggers that activate after a predetermined number of clock cycles~\cite{Kuo2019Time_Related}, data-driven triggers that monitor for specific bit sequences, and execution-based triggers that activate during particular system operations. 

Analog triggers employ circuits that monitor physical parameters such as voltage, current, or temperature. For example, A2 Trojans~\cite{yang_a2_2016} activate when specific voltage or current thresholds are exceeded. While these analog triggers are more difficult to detect than digital ones, they typically still rely on absolute threshold violations that can be identified through comprehensive testing across different operating conditions.

\subsection{Temperature Sensor Vulnerabilities}
\label{subsec:temp}

Recent research has revealed that temperature sensors are vulnerable to electromagnetic interference (EMI) attacks that can precisely manipulate sensor readings~\cite{tu2019trick}. By exploiting the rectification effect in sensor amplifiers, attackers can induce controlled DC voltage offsets through strategic EMI injection. The attack surface is particularly concerning, with demonstrated effectiveness at distances up to 16 meters through exposed wires and PCB traces~\cite{tu2019trick}.

These findings establish several key capabilities: precise control over \textit{induced voltage offsets with a linear relationship to EMI power}, stealthy manipulation that bypasses traditional detection mechanisms, and significant attack range without requiring physical access~\cite{tu2019trick, wu2018characterization}. Our work leverages these vulnerabilities to create a novel hardware Trojan triggering mechanism that monitors the rate of change in sensor measurements rather than absolute values. By leveraging these proven EMI injection techniques to induce controlled rates of change, we demonstrate how sophisticated hardware Trojans can remain dormant yet responsive to precise environmental manipulations while bypassing existing defenses based on redundant sensing and sensor fusion.

\section{Basics of Solar Inverter}
\label{sec:basics}

% This section provides the foundational understanding of solar inverters essential for comprehending our attack.
Solar inverters employ PWM signals as the primary mechanism for power regulation~\cite{kjaer_review_2005}. PWM varies the duty cycle of the signal to control power delivery. The duty cycle is defined as the ratio between the time the signal is on, $T_{on}$, and the time the signal is off, $T_{off}$: $ D = {T_{on}}/{(T_{on} + T_{off})} $

\subsection{Clamped Fly-back DC-DC Converter}
\label{subsec:flyback_dcdc}

\begin{figure}[t!]
    \centering
    % \vspace{-8pt}
    \includegraphics[width=\linewidth,clip, trim=0 12 0 18]{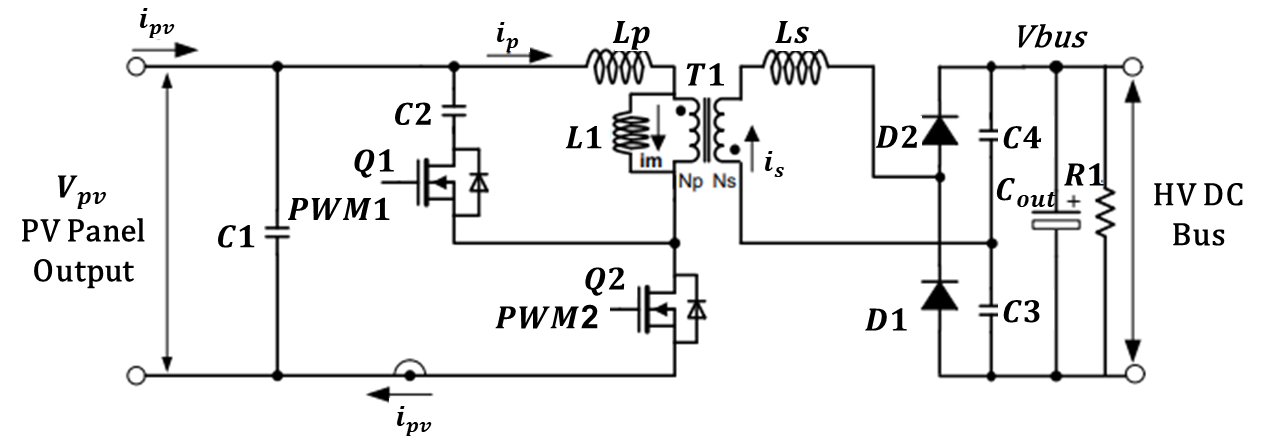}
    \caption{Clamped fly-back DC-DC converter circuit.}
    \label{fig:DCDC}
    % \vspace{-10pt}
\end{figure}

The clamped fly-back DC-DC converter converts the variable DC input into a regulated DC voltage through an active clamping topology shown in Fig.~\ref{fig:DCDC}~\cite{solar_micro_inverter}. The operation involves two complementary PWM signals controlling the main switch (MOSFET \textit{Q2}) and clamping switch (MOSFET \textit{Q1}). When \textit{PWM2} is high, \textit{Q2} conducts, storing energy in transformer \textit{L1}'s magnetic field. During this period, diode \textit{D1} is reverse-biased, preventing current flow to the output. When \textit{PWM2} transitions low and \textit{PWM1} high, \textit{Q2} turns off and \textit{Q1} conducts, allowing the stored energy to transfer through \textit{C2} and the transformer's secondary winding. This energy transfer process generates a boosted DC voltage across the output capacitor \textit{C$_{out}$}. 
The active clamp circuit (\textit{Q1} and \textit{C2}) serves two main functions: recovering energy from the transformer's leakage inductance and suppressing voltage spikes that could damage the main switch~\cite{voltagespike}.
The DC/DC stage is controlled by the MPPT algorithm, which dynamically changes the PWM signals based on the current and voltage sensed from the PV panel. \toremove{The duty cycle changes the amount of energy transferred per cycle in the converter, thereby changing the output voltage and current of the DC-DC to keep the operation at MPPT. 
}

\subsection{DC-AC Inverter}
\label{subsec:dcac_inverter}
The DC-AC inverter~(Fig.~\ref{fig:DCAC}) employs an H-bridge configuration to generate a grid-synchronized AC waveform through two alternating current paths controlled by PWM signals~\cite{solar_micro_inverter}. During the positive half-cycle, current flows from \textit{Vbus} through MOSFET \textit{Q5}, \textit{L4}, \textit{V1}, \textit{L5}, \textit{L3}, and \textit{Q4} (which rapidly switches for power regulation). For the negative half-cycle, current flows through MOSFET \textit{Q6}, \textit{L5}, \textit{V1}, \textit{L4}, \textit{L2}, and \textit{Q3}. These complementary paths enable bidirectional current flow necessary for AC generation. The PWM duty ratios precisely control the energy delivery in each path, while inductors \textit{L2}\textit{-}\textit{L5} and their associated components form an \textit{LCL} filter that reduces harmonic distortion in the output waveform~\cite{poongothai2018design}.

\subsection{Integrated Sensing Systems}
Solar inverters incorporate various sensors for monitoring and protection, including temperature sensors, hall-effect sensors for current measurement, and voltage sensors across critical circuit nodes~\cite{rao2020applications, freitas2022design, meribout2022sensor}. 

This paper focuses on temperature sensing systems, which typically employ sensors like the LM61CIM that produce a voltage output proportional to temperature~\cite{lm61_datasheet}. While we demonstrate our attack using temperature sensors, similar principles could be applied to other classes of sensors that share similar signal conditioning processes due to their inherent susceptibility to environmental manipulations such as electromagnetic interference or magnetic fields~\cite{barua_hall_2020, dan2022novel}.

\section{Threat Model}
\label{sec:threat_model}

\subsection{Attack Scenario and Actors}

Our threat model involves two distinct but potentially coordinated actors: 1) a malicious entity in the manufacturing supply chain who introduces hardware modifications, and 2) field operators who trigger the implanted circuit through environmental manipulation. This separation of roles aligns with established supply chain attack patterns~\cite{ sharma2022complexity, sheng2024pager}.

The supply chain actor could be a rogue employee at a semiconductor foundry, a third-party component manufacturer, or an infiltrated testing facility capable of introducing the minimal 14 $\mu m^2$ hardware modification. The field operator possesses the means to manipulate environmental conditions around targeted inverters through EMI~\cite{tu2019trick, wu2018characterization}.

Critically, these actors need not be the same entity. The supply chain compromise could be conducted by state-sponsored groups sharing targeting information with field operators, a pattern observed in real-world attacks~\cite{guardian_2024, sheng2024pager}.

\begin{figure}[t!]
    \centering
    % \vspace{-8pt}
    \includegraphics[width=\linewidth,clip, trim=0 17 0 13]{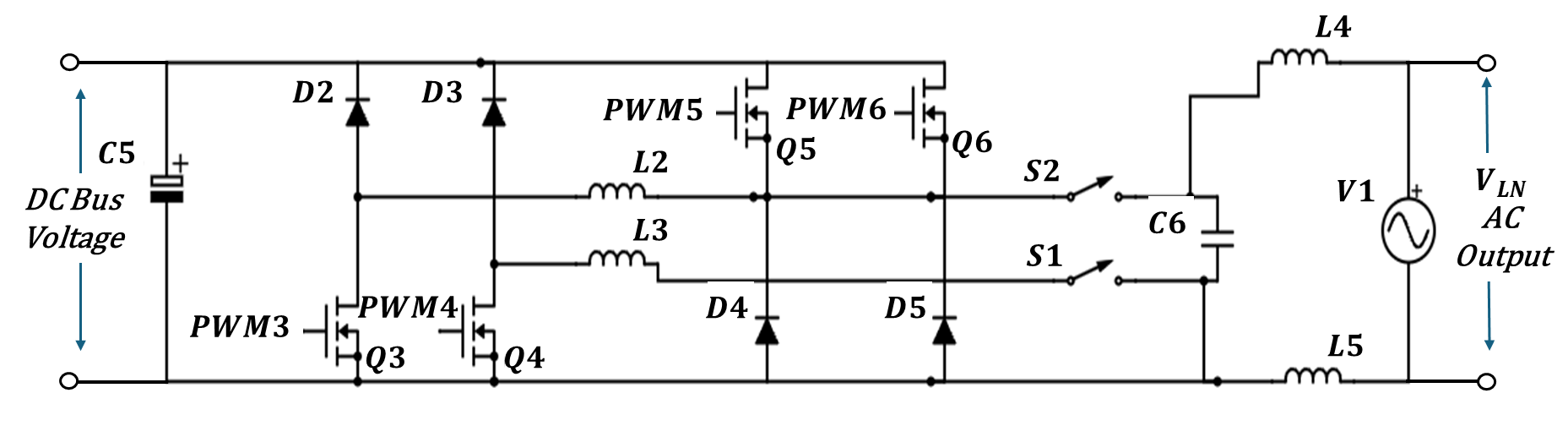}
    \caption{H-bridge DC-AC inverter circuit generating grid-synchronized AC output using PWM-controlled current paths.}

    \label{fig:DCAC}
    % \vspace{-10pt}
\end{figure}

\subsection{Attacker Capabilities and Goals}

The supply chain actor possesses hardware design expertise and manufacturing access. The field operator can generate controlled EMI directed at the temperature sensor to produce specific rates of change~\cite{tu2019trick}. No physical contact with the device is required, as EMI can be effective at distances up to 16 meters~\cite{tu2019trick, wu2018characterization}.
The primary goal is to cause power system disruption while evading detection. Secondary objectives may include demonstrating capability, creating distrust in renewable technologies, or establishing persistent access. The rate-based triggering mechanism can precisely control the attack timing, allowing for strategic activation during high-load periods.

\subsection{Attack Vectors and Impact}
ERM exploits sensor-based protection systems by monitoring capacitor charging rates in environmental sensor circuits rather than absolute thresholds. By applying precisely modulated EMI at specific frequencies, attackers can induce controlled rate-of-change patterns without generating suspicious readings, requiring no physical access.

Once triggered, the malicious circuit manipulates \textit{PWM1} in the DC-DC stage, compromising the active clamping mechanism and creating voltage instabilities. In the DC-AC stage, it disrupts grid synchronization by altering H-bridge switching patterns, ultimately causing cascading failures through grid reactive power-voltage interactions.

\subsection{Defender Assumptions and Limitations}
\label{sec:assumptions}
Defenders implement standard hardware security measures, including redundant sensors, sensor fusion, and anomaly detection. Manufacturing tests validate component resilience across operational extremes.
These defenses prove ineffective against ERM attacks because: 1) Redundant sensor architectures fail because ERM targets the rate of change in a single sensor through EMI. Whether systems use the compromised sensor data or discard it, the attack succeeds as it only needs to trigger the malicious logic through one sensor while keeping readings within normal parameters; 2) Sensor fusion systems fail because ERM operates independently of sensor data processing; and 3) Traditional test-time defenses cannot identify ERM attacks as they monitor rates within normal operating ranges.

\section{ERM Triggered Hardware Trojan}
\label{sec:erm}

\subsection{Novel Triggering Mechanism}
To understand ERM's novelty, we examine how it differs from existing trigger mechanisms:

\subsubsection{Traditional Triggering Approaches and Their Limitations}
Digital triggers rely on specific patterns or sequences in the digital domain, such as time-based counters or data-driven conditions~\cite{naveenkumar2021survey}. Analog triggers monitor physical parameters like voltage or temperature and activate when predefined thresholds are exceeded. External triggers respond to external stimuli, such as radio frequency signals or network commands.

Although these methods have been extensively studied and have well-known detection techniques~\cite{tehranipoor_survey_2010,  xiao_hardware_2016, naveenkumar2021survey}, they often exhibit detectable patterns during testing or side-channel analysis. Digital triggers can be revealed through exhaustive state testing, analog triggers through exposure to varied environmental conditions, and external triggers by analyzing anomalies in electromagnetic emissions or network traffic.

\subsubsection{The ERM Approach}
ERM differs by monitoring the \emph{rate of change} in environmental parameters rather than fixed values or thresholds. It activates when environmental conditions change rapidly, offering advantages over traditional triggering mechanisms.
A key weakness in hardware security testing is that systems are typically evaluated at different temperatures and environmental conditions, but their behavior under specific rates of change is not analyzed. This creates a fundamental blind spot in testing protocols.

ERM triggers through indirect sensor manipulation rather than direct interference with system logic or communication pathways. Unlike analog Trojans such as A2~\cite{yang_a2_2016}, which accumulate charge through repeated transitions in specific wires until reaching an activation threshold, ERM monitors subtle variations in \emph{sensor outputs} and responds to \emph{environmental changes} that do not require \textit{local code execution}. This makes ERM particularly effective against air-gapped or security-hardened systems, where traditional triggers are impractical.

Additionally, ERM allows precise control over activation timing by adjusting local environmental change rates. Its reliance on rate-based activation also reduces the risk of accidental triggers, as natural environmental changes rarely match the exact conditions required for activation.

\subsection{ERM Design}
ERM leverages existing system sensors to detect specific rates of change as a triggering mechanism. 
One of the key innovations in our design is translating rate changes into a controlled triggering signal.
The design addresses key requirements: (1) minimal area, (2) low power consumption, (3) negligible timing impact, and (4) standard cell compatibility.
In our implementation, we exploit the existing temperature sensing circuitry in a Texas Instruments grid-tied solar inverter kit (Part\# TMDSSOLARUINVKIT with a C2000 MCU~\cite{solar_micro_inverter}). 
% \maybe{This demonstrates how ERM can be integrated into commercial power electronics systems.}

\begin{figure*}[t!]
    \centering
    % \vspace{-8pt}
    \includegraphics[width=0.99\linewidth, clip, trim=0 17 0 13]{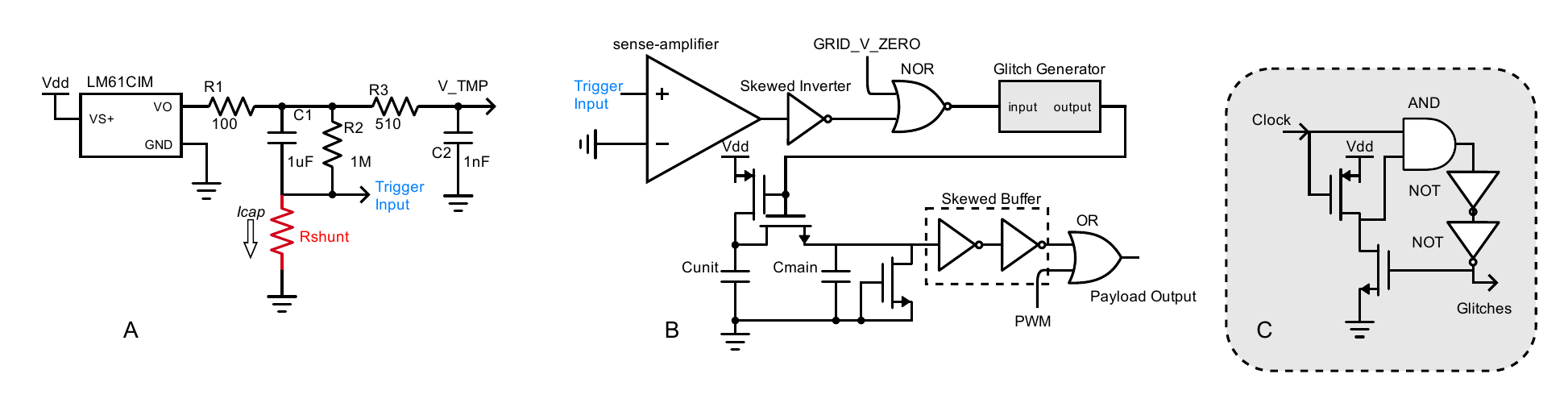}

    \caption{(A) Temperature sensing circuit. (B) Trojan trigger logic with sense amplifier, timing control, \& payload interface. (C) Glitch generator circuit.}

    % \vspace{-8pt}
\label{fig:complete_trojan_design}

\end{figure*}

\subsubsection{General Design Principles}
Fig.~\ref{fig:complete_trojan_design} (A) shows the temperature sampling circuit of our target system, which includes the LM61CIM temperature sensor~\cite{lm61_datasheet} and its associated filtering components. The sensor provides a voltage output proportional to temperature:
\begin{equation}
   V_{TMP} = \alpha T + \beta
\end{equation}
where \( T \) is the ambient temperature in Celsius, and $\alpha=0.01V/^{\circ}C$ and $\beta=0.6V$ are sensor-specific constants~\cite{lm61_datasheet}. 
For any such sensor whose output varies with environmental conditions (temperature, magnetic field, light), the rate of change can be detected by measuring the current through the sensor's output capacitor:
\begin{equation}
   I_{cap} = C \times {dV}/{dt}
\end{equation}
where $dV/dt$ is the rate of change of the sensor's output voltage. Due to the linear relationship between $V_{TMP}$ and temperature:
\begin{equation}
   {dV}/{dt} = \alpha \times {dT}/{dt}
\end{equation}
This relationship enables us to detect specific change rates through the current measurement of the capacitor.

\subsubsection{Circuit Implementation Considerations}
As shown in Fig.~\ref{fig:complete_trojan_design} (A), our design presents three possible approaches for implementing current sensing through shunt resistor placement. 
Using $R_1$ provides the most significant signal amplitude but introduces non-linearity in the measurement. Alternatively, using $R_3$ maintains measurement linearity but results in the smallest signal amplitude. A third option involves introducing a dedicated shunt resistor between $C_1$ and ground, which offers both linearity and sufficient signal amplitude.
We can replace the shunt resistor with the  $R_{ds(on)}$ of a transistor, which is the on-state drain-to-source resistance, and it will significantly reduce the size of the modified parts.
Additionally, the attacker can easily adjust the $R_{ds(on)}$ value according to system constraints and their specific requirements by modifying transistor parameters.

\subsection{Trojan Activation and Payload}
\label{subsec:trojan_activation_payload}

The subtle current generated by sensor reading changes is first processed by a high-precision sense amplifier.
The amplified signal is then passed to a trigger and payload activation circuit, which converts detected environmental rate changes into controlled Trojan activation. 
As shown in Fig.~\ref{fig:complete_trojan_design} (B), the amplifier’s output drives a three-stage trigger system: (1) skewed inverter threshold detection, (2) charge pump timing control, and (3) payload interface.

\subsubsection{Skewed Inverter Threshold}
The skewed inverter's output follows a nonlinear transfer function:
$    V_{out-inv} = f(V_{in}) $, 
where $f(V_{in})$ represents the inverter's nonlinear response to input voltage. The inverter's switching threshold is precisely controlled by adjusting the threshold voltage $V_{th}$ of both \textit{NMOS} and \textit{PMOS} gates. During HT fabrication, this threshold is set within a $0.2V$ to $1V$ range through various techniques including doping concentration modification, work function tuning, and body biasing. These adjustments \textit{ensure the inverter triggers only when the rate of environmental change matches our desired attack conditions}.

When the sense amplifier output stays below this threshold, the inverter holds a logic high. The inverter switches low once the sensor slope rises enough to exceed the threshold. 
The \textit{NOR} gate combines this threshold-dependent signal with GRID\_V\_ZERO, a $60Hz$ periodic signal derived from the inverter's grid synchronization circuitry. The \textit{NOR} gate's output feeds into a glitch generator~(Fig.~\ref{fig:complete_trojan_design} (C)) that produces a precise glitch at each rising edge of its input signal.
% This transition is combined with the 60\,Hz GRID\_V\_ZERO reference in a NOR gate, whose output triggers a glitch generator (Fig.~\ref{fig:complete_trojan_design} (C)). 

\subsubsection{Charge Pump Timing}
The charge pump circuit, activated by these glitches, serves as the core timing control mechanism. It consists of a \textit{PMOS} transistor, an \textit{NMOS} transistor, and two carefully sized capacitors: $C_{unit}$ and $C_{main}$.
Each generated glitch causes $C_{unit}$ to transfer a small, controlled amount of charge to $C_{main}$. 
As these pulses accumulate, $C_{main}$ gradually charges up to a set activation voltage.
This gradual accumulation of charge creates a time delay between initial trigger detection and payload activation, enhancing the attack's stealth.
To prevent detection during inactive periods, we incorporate shunt transistor with shorted gate and source connections, providing a controlled leakage path that allows $C_{main}$'s voltage to decay gradually to zero when the triggering condition is removed.
Consequently, the Trojan remains invisible under typical operating conditions.

\subsubsection{Payload Override}
$C_{main}$'s voltage is continuously monitored by a skewed buffer,  designed with similar threshold characteristics as the input stage's skewed inverter. 
When $C_{main}$ reaches its threshold, the skewed buffer detects this voltage and signals the Trojan’s payload to engage. In our design, the payload overrides the gate driver, for example, for the clamping transistor in the DC-DC boost stage. Under normal conditions, the microcontroller’s PWM signal regulates the active clamp transistor. The Trojan, upon activation, overrides this signal, forcing PWM into a constant logic high or low state.  This manipulation prevents the clamping circuit from recovering energy stored in the transformer's leakage inductance, initiating our intended disruption of the power conversion process. 

In contrast to A2 Trojans~\cite{yang_a2_2016}, which accumulate charges from specific digital activity and may require a malicious software routine for Trojan triggering, the ERM triggering mechanism operates based on the rate of change in sensor voltage. 
This activation requires \textit{neither specific software/code execution nor abnormal sensor readings}, making it inherently more stealthy. 

\subsubsection{Trojan Sensitivity Control}
The Trojan's sensitivity to rate changes is controlled through two key design parameters: $C_{main}$'s capacitance and the sense amplifier's gain $A_v$. The capacitor size $C_{main}$ determines the charging time constant and, thus, the delay between detection and activation. The amplifier gain $A_v$ sets the circuit's sensitivity to rate changes. These parameters are carefully balanced to ensure reliable triggering under attacker-induced conditions while preventing false activation from natural environmental variations. In our implementation, these parameters are tuned to respond to temperature changes exceeding 0.1$^{\circ}$C per second, a rate that indicates deliberate manipulation rather than natural variation.

\subsubsection{Trojan Trigger Timing Control} The Trojan's trigger timing is controlled by five core parameters: (1) $C_{main}$ capacitance, (2) $R_{ds(on)}$ of the shunt transistor,, (3) NOR gate input frequency (glitch generator frequency), (4) leakage transistor characteristics using our 45nm process library, this transistor exhibits high resistance ($>100M\Omega$) at low $V_{ds}$ with saturation in leakage current at higher $V_{ds}$, and (5) skewed buffer threshold voltage. These parameters offer flexibility in attack design: $C_{main}$'s size directly affects trigger duration but impacts circuit footprint; higher $R_{ds(on)}$ can compensate for limited sense amplifier gain; and while our implementation uses the 60Hz GRID\_V\_ZERO signal, the glitch generator frequency could be adjusted to modify timing. Furthermore, trigger duration can be extended by reducing leakage current or increasing the skewed buffer's threshold voltage. In our implementation, we achieve a 5-second trigger delay using a 0.7V skewed buffer threshold.

\subsubsection{Stealth and Low Power}
The entire trigger and payload circuit draws negligible current during both active and inactive states. Its small footprint, similar to a single standard cell, allows it to blend into filler regions. The circuit remains quiescent when environmental changes occur at typical speeds. Only a deliberately injected burst of EMI or a sudden rapid shift in temperature triggers the sense amplifier, gradually charging $C_{main}$ until the payload activates. 

\begin{figure*}[t!]
    \centering
    \includegraphics[width=0.9\linewidth]{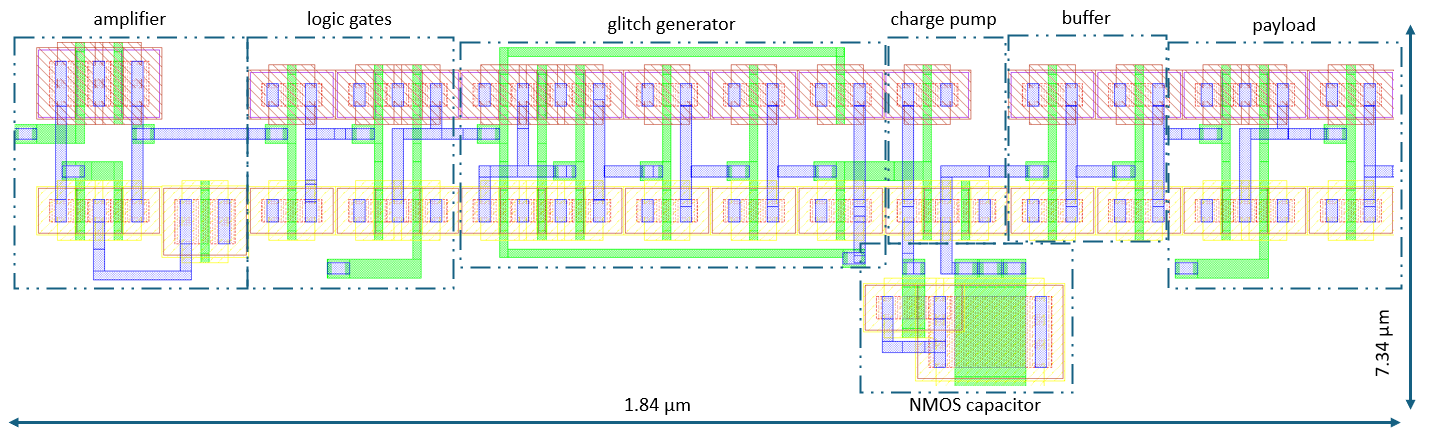}
    % \vspace{-9.9pt}
    \caption{HT layout in Cadence Virtuoso at 45\,nm, occupying \(<14~\mu\text{m}^2\) within inverter control logic.}
    \label{fig:HT_layout_modified}
    % \vspace{-10pt}
\end{figure*}

\subsection{Integration and Minimization of Detection}
\label{sec:integration_minimization_detection}

We integrated the Trojan into the inverter’s temperature sensing and PWM control circuits so it would blend with existing hardware rather than appear as an isolated macro. As shown in Fig.~\ref{fig:HT_layout_modified}, our Cadence Virtuoso design at the 45\,nm node~\cite{gpdk045_library} occupies less than \(14~\mu{m}^2\) (less than 0.00005\% of the chip size). This footprint is comparable to a single standard cell, allowing the Trojan to fit unnoticed among filler cells. 
The ERM Trojan is meticulously designed to evade existing hardware Trojan detection methodologies, which often fall short due to ERM's unique characteristics:

\subsubsection{Side-Channel Analysis Evasion}
    \begin{itemize}[leftmargin=10pt]
        \item \textit{Power Analysis:} The Trojan's CMOS structure yields minimal static power. Its dynamic power (from $P=0.5 \times CV^2 \times \text{frequency}$) is below 1~nanoWatt, considering pF-range capacitance ($C_{\text{main}} + C_{\text{unit}}$), a 1.2V supply, and a 50\,Hz glitch generator. This minuscule power signature is exceptionally difficult to detect against the solar inverter's larger, variable power consumption due to MPPT operations \cite{bhatta2024machine}.
        \item \textit{Timing Analysis:} The payload's timing impact is negligible, adding only an OR gate delay (114\,ps with a 10\,fF load). The inverter's low-speed MCU controller makes traditional clock-sweeping inapplicable \cite{xiao2013clock}. Advanced methods like LASCA \cite{vakil2020lasca}, while sensitive to minor delays, struggle with PVT variations and require precise, stable testing conditions under which ERM, by design (triggering on \textit{rate of change}), remains inactive. 

    \end{itemize}

\subsubsection{Optical Inspection Evasion} Though optical inspection can detect manufacturing-time Trojans \cite{bhasin2015survey}, its high cost, time requirements, and complications from varied mask sets limit its practicality for large IC volumes. The ERM Trojan's tiny $14~\mu\text{m}^2$ footprint, comparable to standard cells, makes it unlikely to be flagged as an anomaly.

\subsubsection{Testing-Based Detection Evasion} Standard functional correctness tests are unlikely to activate ERM, as it triggers on specific environmental parameter \textit{rates of change} (e.g., $>$0.1$^{\circ}$C/sec), not static abnormal values. Replicating these precise dynamic conditions during rare-event testing is difficult without prior knowledge of the trigger \cite{chakraborty2009mero}. 

\subsubsection{Run-Time Detection Evasion} ERM remains dormant under normal operating and factory testing conditions, eluding run-time monitors looking for abnormal signals. The inverter's dynamic environment (temperature, load changes) can further mask subtle cues \cite{bloom2009support}. Moreover, the overhead of effective run-time detection conflicts with the design goals of energy-sensitive solar inverters.

\section{Trojan Integration and Component-Level Analysis}
\label{sec:implementation}

\subsection{Integration Strategy}
We integrate our Trojan at two critical control points in the solar inverter's power conversion stages:

\subsubsection{DC-DC Converter Integration}
The DC-DC stage converts variable input from the solar panel to stable DC voltage through a flyback converter with active clamping~(ref. to Fig.~\ref{fig:DCDC} in Sec.~\ref{subsec:flyback_dcdc}). The relationship between the output voltage ($V_{out}$) and the solar panel's input voltage ($V_{in}$) follows:
\begin{equation}
   V_{out} = V_{in} \times {N_s}/{N_p} \times {1}/({1-D_{main}})
   \label{eqn:flyback_output}
\end{equation}
where $N_s/N_p$ is the transformer's turns ratio, and $D_{main}$ is the duty ratio of the main transistor.

\toremove{
The solar inverter implements active clamping functionality on top of the conventional flyback inverter to suppress spike voltage when the primary MOSFET is cut off and to recover energy from the transformer's leakage inductance.

Our design targets \textit{PWM1}, which controls the active clamping transistor. The main transistor is controlled via \textit{PWM2}, which dictates the on-off cycles at high frequency to stabilize the voltage output. In typical operation, the microcontroller modulates this signal to regulate the voltage, maintaining the solar inverter's maximum power point. \textit{PWM1} is complementary to \textit{PWM2}, controlling the clamping transistor crucial for efficient energy transfer and minimal power loss. The clamping transistor also protects the main transistor by absorbing voltage spikes during switching transitions.}
Our design targets \textit{PWM1}, which controls the active clamping transistor. In normal operation, the microcontroller modulates this signal to maintain optimal energy transfer from leakage inductance while protecting the main transistor from voltage spikes. This signal is particularly effective as an attack vector because it directly impacts both efficiency and protection mechanisms without immediately triggering fault detection systems.

\subsubsection{DC-AC Inverter Integration}
The DC-AC inverter relies on high-frequency PWM signals to control the switching of the MOSFETs in the H-bridge configuration~(ref. to Fig.~\ref{fig:DCAC} in Sec.~\ref{subsec:dcac_inverter}). During positive half-cycle generation, current flows from $V_{bus}$ through \textit{Q5}, \textit{L4}, \textit{V1}, \textit{L5}, \textit{L3}, and finally \textit{Q4} (which rapidly switches for power regulation). For the negative half-cycle, current flows through Q6, \textit{L5}, \textit{V1}, \textit{L4}, \textit{L2}, and \textit{Q3}. The combination of these switching patterns produces a bipolar output that is filtered to create a clean AC waveform synchronized with the grid's frequency and phase~\cite{solar_micro_inverter}.

Our integration targets \textit{PWM5}, which works in conjunction with \textit{PWM6} to control the positive \& negative half-cycles, respectively. \textit{PWM6} operates as a complementary signal to \textit{PWM5}. Additionally, \textit{PWM3} and \textit{PWM4} adjust their switching to modulate the AC output amplitude. This arrangement is critical in grid-tied solar inverters, as any disruption in the inverter's output can lead to power quality issues or synchronization problems with the grid.

\subsection{Attack Scenarios and Mechanisms}

\subsubsection{DC-DC Converter Attack Scenarios}
\label{subsubsec:ltspice_dcdc_attack}
We implement two distinct attack vectors with different severity profiles:

\textbf{\textit{PWM1} High State Attack:} When \textit{PWM1} is locked in a high state, the clamping transistor remains continuously conducting. This forces energy stored in the transformer's leakage inductance to be dissipated through the clamping circuit rather than transferred to the output. The continuous current flow creates thermal stress on both the clamping transistor and capacitor while gradually reducing DC-DC converter's output voltage. The voltage drop is gradual enough to avoid triggering immediate overcurrent protection while still significantly impacting the converter's efficiency.

\textbf{\textit{PWM1} Low State Attack:} 
A more severe scenario occurs when \textit{PWM1} is locked in a low state, completely disabling the active clamping functionality. Without the clamping circuit's protection, the energy stored in the transformer's leakage inductance has no controlled path for dissipation when the main switch turns off. This leads to large voltage spikes across the main switch due to the rapid change in current through the leakage inductance $V = L \times di/dt$~\cite{voltagespike}. These voltage spikes can exceed the voltage rating of the photovoltaic full-bridge driver chip (SM72295MA)~\cite{sm72295_datasheet}, leading to catastrophic failure. Additionally, the voltage spikes create resonant oscillations with parasitic capacitances in the circuit, generating high-frequency noise that can further stress circuit components.

\subsubsection{DC-AC Inverter Attack Mechanism}
\label{subsubsec:ltspice_dcac_attack}
The attack manipulates the H-bridge configuration through \textit{PWM5}:

\textbf{\textit{PWM5} High State Attack:} Locking \textit{PWM5} in a high state forces \textit{Q5} to remain continuously on, disrupting the H-bridge switching pattern. During positive half-cycles, the operation appears relatively normal as \textit{Q5} would typically be on. However, during negative half-cycles, the constant on-state of \textit{Q5} creates a path that reduces the achievable voltage differential across the output. When \textit{Q6} attempts to conduct to generate the negative half-cycle, the presence of the conducting \textit{Q5} creates a partial short-circuit path that diverts current from the intended output path, causing a significant reduction in the negative half-cycle amplitude.

The asymmetric conduction fundamentally alters the output voltage waveform, reducing negative peaks and decreasing fundamental frequency components. The resulting waveform distortion introduces significant harmonic content into the output. The phase relationship distortion makes grid synchronization difficult, while the asymmetric voltage output affects power quality. Under ideal loading conditions and grid connection characteristics, this can result in approximately 50\% reduction in negative half-cycle amplitude. 

\subsubsection{Attack Coordination}
The activation of these attacks is precisely controlled through the charge pump circuit. When environmental sensors detect a rate of change exceeding the predetermined threshold (0.1$^{\circ}$C per second for temperature-based triggering), the charge pump accumulates charge through controlled pulses synchronized with the grid frequency using the GRID\_V\_ZERO signal. This gradual activation approach ensures a smooth transition into the attack state, avoiding transient spikes that might trigger protection mechanisms while maximizing the attack's impact.

\subsection{Component-Level Simulation and Analysis}
\label{subsec:component_level_sim}
We validate our attack through comprehensive component-level simulations in LTSpice~\cite{ltspice}, demonstrating technical feasibility and potential impact on system integrity.

\begin{figure*}[t!]
    \centering
    \includegraphics[width=0.859\linewidth]{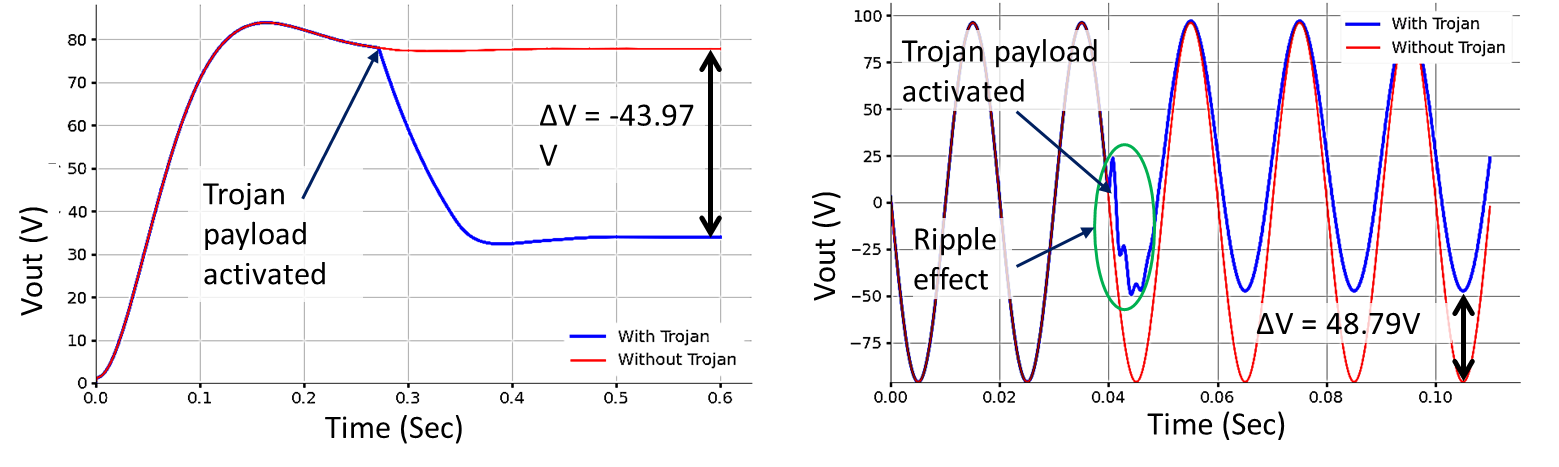}
    \caption{Output voltage of DC-DC converter (left) and DC-AC inverter (right) before and after Trojan activation, showing significant voltage degradation.}

    % \vspace{-10.0pt}
\label{fig:dcdc_and_dcac_result_ltspice}
\end{figure*}

\subsubsection{DC-DC Converter with HT}
\label{subsubsec:ltspice_dcdc_attack_demo}
Our LTSpice model integrates HT into the clamped flyback DC-DC converter circuit (Fig.~\ref{fig:DCDC}), allowing us to analyze the attack. 

\textbf{Simulation Results:} Fig.~\ref{fig:dcdc_and_dcac_result_ltspice} (left)
demonstrates the converter's behavior before and after attack activation at $0.27s$. Initially, the DC-DC converter operates normally, boosting and regulating the input voltage as described in Eq.~\ref{eqn:flyback_output}. 
At $0.27s$, the Trojan detects the target \textit{dV/dt} from the sensor's output and activates its payload. The payload locks the clamping transistor’s gate signal (\textit{PWM1}) to a high state, disrupting the clamping mechanism.
The immediate impact is evident in the rapid decline in output voltage from $77.77V$ to $33.80V$, representing a $56.53\%$ reduction in power efficiency. \extra{(Detailed explanation of the calculation can be found in the Appx.~\ref{appx:efficiency_degradation}).}
Although the output voltage eventually stabilizes, it remains at a reduced level, demonstrating the Trojan's ability to degrade the converter’s performance significantly.
In a real-world scenario, the continuous current flow through the clamping circuit can cause significant heat dissipation in the clamping transistor and capacitor, exceeding safe thermal limits.
\toremove{demonstrates the converter's behavior before and under attack. Initially, the DC-DC converter operates normally, boosting and regulating the input voltage as described in Eq.~\ref{eqn:flyback_output}. 
At 0.27s, the Trojan detects the target \textit{dV/dt} from the sensor's output and activates its payload. The payload locks the clamping transistor’s gate signal (\textit{PWM1}) to a high state, disrupting the clamping mechanism.
The immediate impact is evident in the rapid decline in the converter’s output voltage from its nominal 77.77V to 33.80V, representing a 56.53\% reduction in power efficiency (Detailed explanation of the calculation can be found in the Appx.~\ref{appx:efficiency_degradation}). 
Although the output voltage eventually stabilizes, it remains at a reduced level, demonstrating the Trojan's ability to degrade the converter’s performance significantly.
In a real-world scenario, the continuous current flow through the clamping circuit can cause significant heat dissipation in the clamping transistor and capacitor, exceeding safe thermal limits. 
}

\subsubsection{DC-AC Inverter with HT}
\label{subsubsec:ltspice_dcac_attack_demo}
The DC-AC inverter, shown in Fig.~\ref{fig:DCAC}, was modeled in LTSpice with the Trojan embedded in its H-bridge configuration. 

\textbf{Simulation Results:} Fig.~\ref{fig:dcdc_and_dcac_result_ltspice} (right) 
depicts the AC voltage output before and after HT activation. The inverter initially produces a symmetric sinusoidal waveform with 192.30V peak-to-peak voltage. At 0.04s, the Trojan activates, locking \textit{PWM5} to a constant high state. Following activation, the inverter exhibits a significant 50.74\% reduction in negative half-cycle amplitude, creating an asymmetric output waveform.

This fault is caused by the locked \textit{Q5} transistor, which creates a partial short-circuit path.
In subsequent negative cycles, regardless of how the other MOSFET \textit{Q6} switches, the MOSFET near \textit{Vbus} \textit{Q5} remains continuously on, disrupting the normal H-bridge switching mechanism. When \textit{Q6} attempts to turn on, the presence of a short-circuit diverts part of the current, resulting in a reduced voltage across the load. 
The Root Mean Square (RMS) voltage without Trojan is $67.97 V$, while it drops to $56.74 V$ with Trojan, representing a reduction of $16.52\%$.
This level of voltage loss may impair the inverter’s ability to meet grid standards for voltage requirements, potentially disrupting grid stability, and causing power and efficiency losses across the entire grid system.
Ripples in the waveform during the immediate post-trigger period reflect abrupt changes in the inductor current of the AC output filter.

\subsubsection{Potential for System-Wide Failure}
The combined effects of voltage ripple, harmonic distortion, and phase shifts increase the likelihood of system-wide failure. The instability introduced by the Trojan in both stages can lead to thermal stress on critical components, potentially resulting in permanent damage as demonstrated in our hardware validation tests (Sec.~\ref{subsec:hardware_implementation}). This risk is amplified in grid-tied configurations where instability in a single inverter can propagate to other connected devices, leading to broader system-wide consequences as shown in our grid-level simulations (Sec.~\ref{subsec:grid_level_impact}).

\section{Evaluation}
\label{sec:evaluation}
We extend our component-level analysis to evaluate system-wide implications of the ERM attack through three complementary approaches: full system simulation using Simulink~\cite{simulink}, hardware-in-the-loop validation, and power grid impact analysis using ETAP~\cite{etap}.

\subsection{Complete Solar Inverter Simulation}
\begin{figure}[t!]
    \centering
    \includegraphics[width=\linewidth]{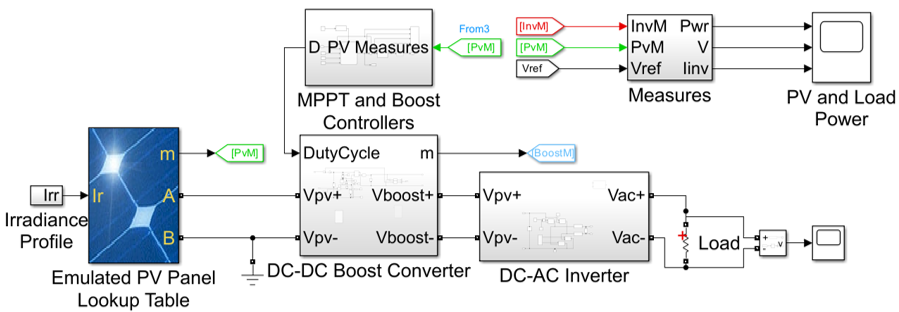}
    \caption{Simulink model of a complete solar inverter system.}
    \label{fig:Simulink_Model}
    % \vspace{-12.0pt}
\end{figure}

Using Simulink, we developed a system-level model (Fig.~\ref{fig:Simulink_Model}) that accurately represents a complete solar inverter with integrated HT. Based on Texas Instruments' reference designs~\cite{solar_micro_inverter}, our model includes an emulated PV panel, DC-DC Active Clamp Flyback Boost Converter, single-phase DC-AC inverter, and MPPT algorithm. 
Unlike the LTSpice simulations with idealized voltage sources, this model employs a PV panel with finite power capacity, creating realistic constraints on system behavior under attack.

\subsubsection{DC-DC Stage Attack Effects}
\label{subec:HT_in_DCDC}
In the DC-DC boost circuit (see Fig.~\ref{fig:Simulink_DC-DC_Converter_Model} in Appx.~\ref{appx:simulink_dcdc}), we replicate our component-level attack by targeting the same clamping transistor control signal (\textit{PWM1}), consistent with our LTSpice analysis (see Sec.~\ref{subsubsec:ltspice_dcdc_attack} and Sec.~\ref{subsubsec:ltspice_dcdc_attack_demo}). 

\begin{figure}[h!]
    \centering
    % \vspace{-5.0pt}
    \includegraphics[width=0.99\linewidth, clip, trim={0 10 0 0}]{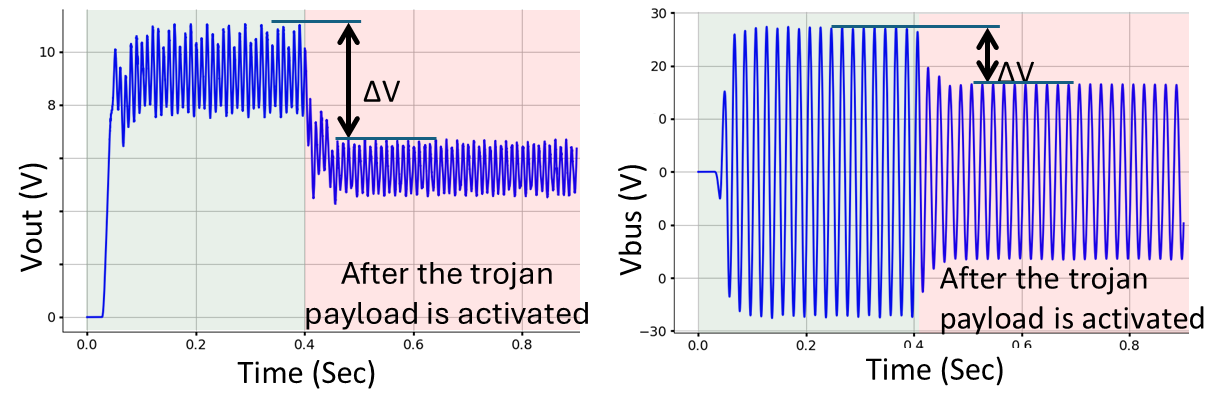}
    \caption{Simulink results showing voltage degradation in (left) DC-DC converter \& (right) DC-AC inverter after HT activation in the DC-DC stage.}
\label{fig:DCDC_and_DCAC_output_with_DCDC_HT_simulink}
    % \vspace{-5.0pt}
\end{figure}

Our simulation results (Fig.~\ref{fig:DCDC_and_DCAC_output_with_DCDC_HT_simulink} (left)) closely align with the component-level analysis: when the Trojan locks the clamping control signal high at $t=0.4s$, we observe a significant output voltage reduction from $11V$ to approximately $6.7V$ ($39\%$ decrease). This degradation occurs through the same mechanism demonstrated in our LTSpice simulations -- disruption of energy recovery from transformer leakage inductance (measured at $2.8 \mu H$). The MPPT algorithm attempts to compensate by modulating the main transistor's duty cycle from $0.65$ to $0.72$, but cannot overcome this fundamental energy loss, matching our earlier findings of compromised efficiency.

The voltage degradation propagates through to the DC-AC stage (Fig.~\ref{fig:DCDC_and_DCAC_output_with_DCDC_HT_simulink} (right)). The $V_{bus}$ voltage drop stems directly from the compromised clamping circuitry, and without boost capability in the DC-AC stage, this degradation manifests as reduced AC output voltage. The overall system efficiency drops by $14.52\%$, validating our component-level predictions of cascading effects through both stages.

\subsubsection{DC-AC Stage Attack Impact}
\label{subsec:HT_in_DCAC}
The DC-AC stage uses an H-bridge configuration with symmetric current paths to generate boosted AC during both half-cycles. This symmetry means that the behavior of one half-cycle is mirrored in the other. During the positive half-cycle, current flows from \textit{V\_{bus}} through \textit{Q5}, \textit{L4}, \textit{L5}, \textit{L3}, and \textit{Q4} to ground. 
Similar to our LTSpice analysis (Sec.~\ref{subsubsec:ltspice_dcac_attack} and Sec.~\ref{subsubsec:ltspice_dcac_attack_demo}) where we targeted \textit{PWM5}, 
We manipulate the control signal that controls the positive half-cycle switching, forcing \textit{Q5}(~\ref{fig:DCAC}) into a continuous conduction state.

\begin{figure}[t!]
    \centering
    \includegraphics[width=0.99\linewidth, clip, trim={0 10 0 0}]{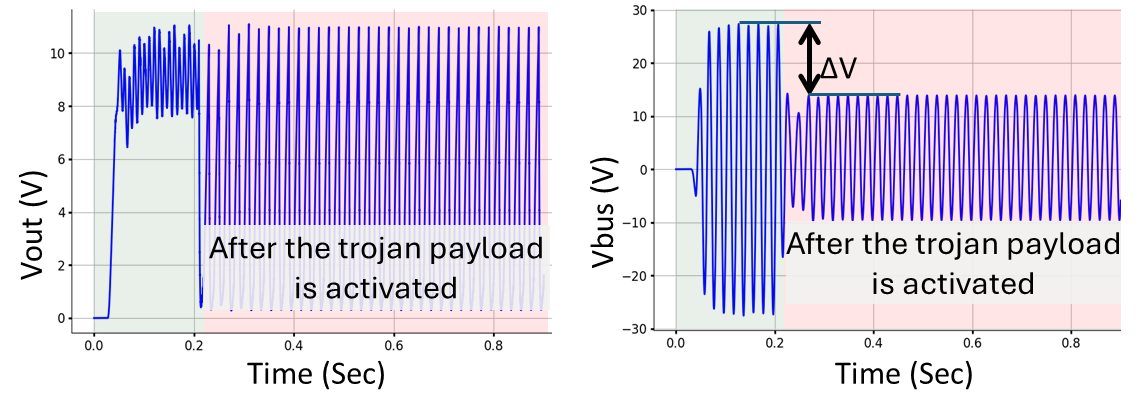}
    \caption{Simulink results showing voltage degradation in (left) DC-DC converter \& (right) DC-AC inverter after HT activation in the DC-AC stage.}
    \label{fig:DCDC_and_DCAC_output_with_DCAC_HT_simulink}
    % \vspace{-10.0pt}
\end{figure}

The attack creates a cascading effect beginning at the DC-DC stage (Fig.~\ref{fig:DCDC_and_DCAC_output_with_DCAC_HT_simulink} (left)). Pre-trigger $V\_{bus}$ voltage maintains approximately $9 V$, but post-activation levels show periodic collapse to $0.3 V$ during switching transitions. This instability stems from the interaction between the locked MOSFET's current path and and the PV panel's limited current sourcing capability (maximum $8A$ in our configuration). The locked transistor in the DC-AC inverter causes current to flow directly to ground, rapidly pulling the Vbus voltage down to near zero during each negative half-cycle. 
During the positive half-cycle, as the transistor turns off, the Vbus voltage can rise for a period of time. This results in an extremely unstable waveform for Vbus.

The final output waveform of the inverter taken at the DC-AC inverter (Fig.~\ref{fig:DCDC_and_DCAC_output_with_DCAC_HT_simulink} (right)) output reveals a system-level effect absent in component-level analysis: bilateral waveform degradation.
Unlike the LTSpice simulations, in which only the negative half-cycle amplitude was affected, the system-level analysis reveals more severe impacts.
This comprehensive waveform distortion stems from the interaction between power stages and source limitations. 
During negative half-cycles, the locked MOSFET creates a low-impedance (short-circuit) path from $V\_{bus}$ to ground, causing a substantial current draw. 
The emulated PV panel, with its finite output capacity, cannot maintain $V\_{bus}$ voltage under the increased load, leading to source collapse and reduced amplitude in both half-cycles. 
The simulation shows peak voltage reduction from $27.16 V$ to $13.9 V$, with RMS voltage dropping from $19.2 V$ to $9.82 V$. This behavior reveals how realistic power source constraints amplify attack impacts beyond what component-level analysis predicts.
\begin{figure}[b!]
    \centering
    % \vspace{-10.0pt}
    \includegraphics[width=0.99\linewidth]{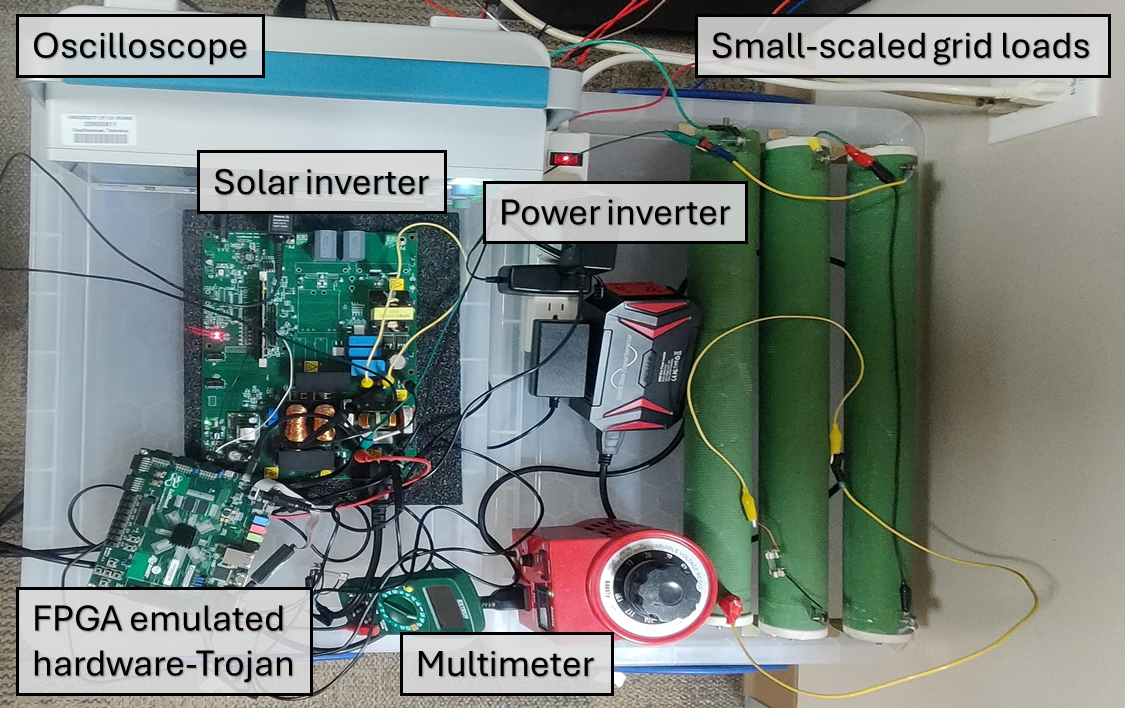}
    \caption{Hardware testbed with TI solar inverter, FPGA-emulated Trojan, and scaled-down grid setup for real-world attack validation.}
    \label{fig:testbed}
\end{figure}

\subsection{Hardware Implementation and Validation}
\label{subsec:hardware_implementation}

To validate our attack's real-world feasibility, we implemented a hardware testbed using a Texas Instruments solar inverter kit (TMDSSOLARUINVKIT) controlled by the C2000 Piccolo-B MCU~\cite{solar_micro_inverter}.
Our experimental setup comprised a scaled-down $140W$ power grid to avoid safety concerns (see Fig.~\ref{fig:testbed}). DC input was simulated through a PSB2400L2 power supply configured to deliver $110V$, $60Hz$ AC power to the grid through three series-connected $160~\Omega $ load resistors.
\subsubsection{FPGA-Based Implementation}
We implement the trigger using an FPGA interfacing with the inverter through PMOD connections~\cite{zedboard_hw_ug}. Since FPGAs cannot directly implement analog circuits, we constructed the analog portion of the hardware Trojan using discrete components on a breadboard. 
The FPGA handles digital logic, generating glitches and controlling the timing of the Trojan's trigger and payload activation. 

The design incorporates synchronized modules for precise attack execution. The trigger detection module employs calibrated I/O thresholds matching the temperature sensor's output characteristics across its operating range. A digital comparator, synchronized to the FPGA's $100MHz$ clock, monitors the sensor output through a dedicated analog-to-digital channel.
The glitch generation module maintained precise synchronization with the $60Hz$ GRID\_V\_ZERO signal, essential for maintaining attack stealth while facilitating charge accumulation. The PWM control module interfaced directly with the inverter's control signals through hardware-implemented PMOD connections, intercepting \textit{PWM1} from the inverter's control card and providing modified signals while maintaining critical timing relationships.

The analog portion is implemented in a breadboard, incorporating a precision op-amp for sensor current amplification, charge pump circuitry, and temperature-compensated threshold detection. A dedicated charge detection module monitored the analog circuit's voltage accumulation through an ADC channel calibrated to the charge pump's $0.2V$ to $1V$ operating range.

\subsubsection{Experimental Setup and Integration}
The experimental setup interfaced the FPGA with the Texas Instruments solar inverter through PMOD connections. Code Composer Studio~\cite{CodeComposerStudio} provided real-time control and monitoring capabilities through its supported ControlSUITE$^{\scriptsize\text{TM}}$ software kit. 

The FPGA design simulated the Trojan's behavior by manipulating PWM signals under specific environmental triggers. 
The trigger's activation caused the charge pump to accumulate voltage over time, eventually modulating the PWM signals in three conditions:

\begin{itemize}
    \item \textit{PWM1} locked at logic low, forcing the active clamping transistor into permanent off-state
\item \textit{PWM1} locked at logic high, forcing the active clamping transistor into permanent on-state
\item \textit{PWM5} locked at logic high, forcing the half-cycle controlled transistor into permanent on-state
\end{itemize}
{

\subsubsection{Experimental Results and Insights}
Our initial test targeted the DC-DC converter stage by locking \textit{PWM1} in a low state. The attack proved immediately catastrophic. The disabled clamping circuit allowed uncontrolled voltage spikes from transformer leakage inductance, leading to rapid failure of the photovoltaic full-bridge driver chip (SM72295MA). The converter output collapsed to $-0.7V$, corresponding to the transformer secondary diode's forward voltage drop.

The failure mode was more severe than predicted by the simulation. The failure of the driver chip, along with subsequent short circuits across the clamp and flyback transistor pins, permanently damaged the solar inverter, despite our attempts at protective measures. Consequently, even after replacing the driver chip, the solar inverter cannot be restored to its original functionality, which prevents further experimentation in subsequent scenarios.
This immediate and catastrophic hardware failure provides key insights:

\textbf{1)} The rapid driver chip failure demonstrates how loss of clamping functionality can exceed component stress tolerances, even in commercial-grade hardware.

\textbf{2)} While simulations showed performance degradation, physical implementation revealed voltage spike propagation occurs too quickly for protection circuits to respond timely.

\textbf{3)} The initial driver chip failure led to subsequent component damage, illustrating how localized failures can cascade through interdependent power conversion stages.

These hardware results validate that real-world component interactions can amplify attack impacts beyond theoretical predictions, creating more severe consequences than simulation alone would suggest.

% These hardware results complement our simulation findings by demonstrating that real-world component interactions can amplify attack impacts beyond predicted levels. The permanent system damage from a single PWM manipulation validates our attack's potential for significant infrastructure disruption, even though it prevented complete experimental validation of all attack scenarios.

\subsection{Grid-Level Impact Analysis}
\label{subsec:grid_level_impact}
% We conducted extensive power system simulations using ETAP~\cite{etap} to demonstrate how device-level vulnerabilities propagate to grid-scale disruption
To evaluate the broader implications of ERM attacks on power grid stability, we conducted extensive power system simulations using ETAP~\cite{etap}. Our analysis demonstrates how device-level vulnerabilities in solar inverters can propagate through distribution networks, potentially triggering cascading failures and widespread system disruption.

\begin{figure}[t!]
    \centering
    \vspace{-5pt}
    \includegraphics[width=\linewidth, trim={0px 40px 0px 0px}]{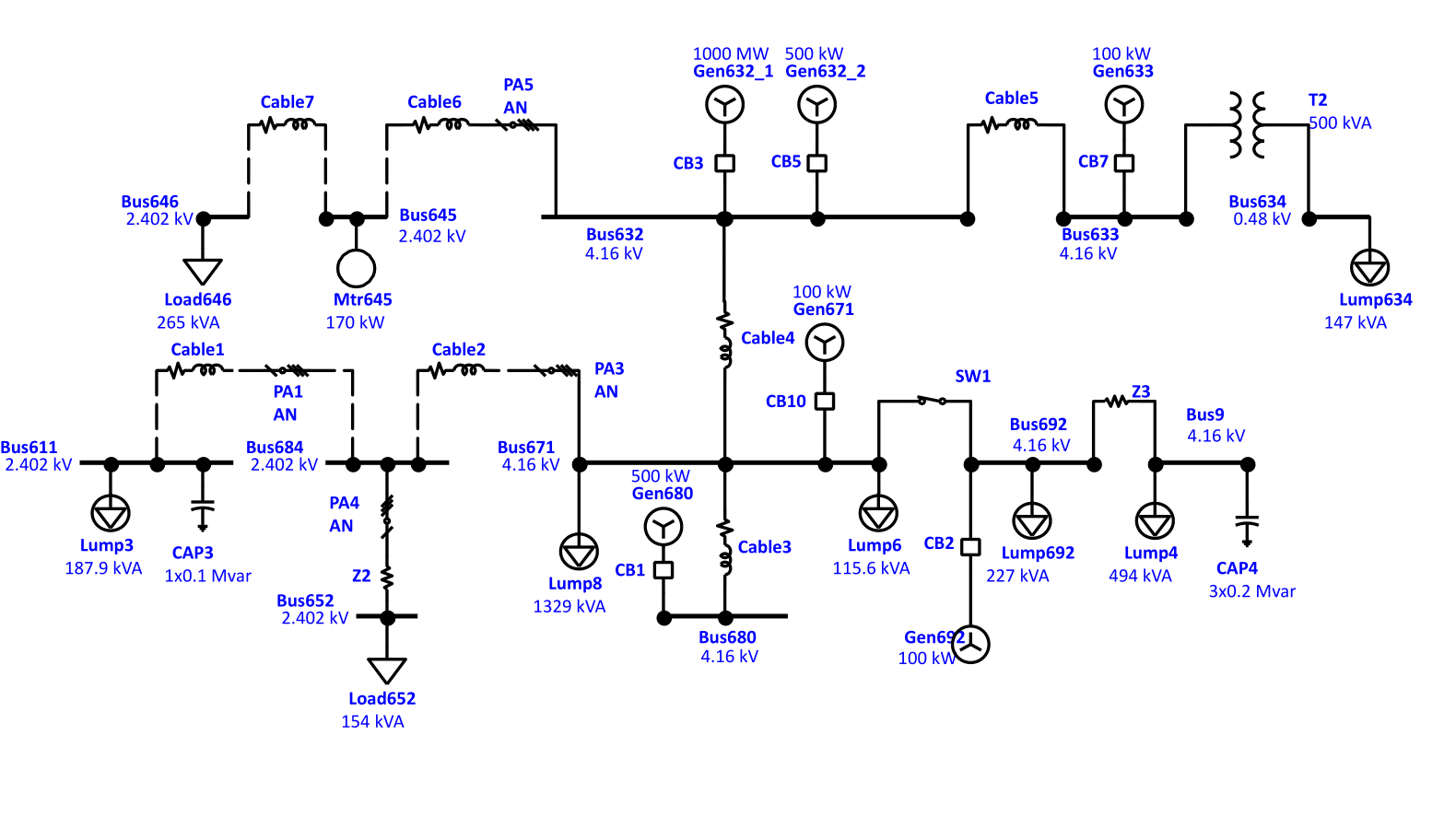}
    \caption{IEEE 13-bus grid with loads and distributed generation.}

    \label{fig:ieee_13}
    \vspace{-10pt}
\end{figure}

\subsubsection{Power Grid Test Environment}
We modeled an IEEE 13 bus distribution test system (Fig.~\ref{fig:ieee_13}), to evaluate attack propagation in a realistic network environment. 
This model was specifically chosen for its representation of typical North American medium-voltage radial distribution systems. 
The test grid comprises approximately $2.6 MW$ and $1.8 MVar$ of distributed loads, equivalent to the power requirements of a medium-sized community or small town. A $1000 MW$ main generator connects to the slack bus, supplemented by distributed generation units ranging from $100 kW$ to $500 kW$.

The target $100 kW$ solar generation unit at bus 692 was chosen based on its electrical distance from the main generator and proximity to critical loads.
The network operates at 4.16 kV primary voltage, incorporating mixed load types (residential, commercial, industrial) and industry-standard protection schemes. 
Network parameters were selected to ensure realistic system behavior. The short circuit capacity at the utility bus is set to $1000 MVA$, providing sufficient fault current for proper protection system operation while maintaining realistic system strength characteristics. Power factor correction is implemented through local capacitor banks, and voltage regulation is achieved through load tap changers on critical feeders. The network includes overhead lines and underground cables with practical X/R ratios, affecting power flow and voltage stability characteristics.

\subsubsection{Attack Manifestation and Initial Response}
Based on our Simulink simulation and hardware-level analysis, we identified three distinct failure modes in grid-connected solar inverters under ERM attack:

I) Catastrophic device failure, demonstrated through the complete failure of the photovoltaic full-bridge driver chip, implemented as an instantaneous generator trip at bus 692.

As shown in Fig.~\ref{fig:ETAP_Voltage_Impact_Fault} (A), when the generator disconnects from the grid, the main grid frequency continuously decreases linearly. This occurs because, during peak demand periods, other generators lack sufficient frequency regulation capacity. In the case of overloading other generators, a power imbalance arises between generation and load, causing the frequency to drop. According to the IEEE 1547 standard~\cite{generation2020ieee}, the grid will shut down when the frequency drops below 59.3 Hz. Therefore, if the generator remains disconnected, it may ultimately lead to a grid-wide blackout.

II) Severe voltage degradation, corresponding to 16.52\% reduction in output RMS voltage (Sec.~\ref{subsec:component_level_sim}), modeled through modified generator voltage regulator characteristics.

The voltage impact attack (Fig.~\ref{fig:ETAP_Voltage_Impact_Fault} (B)) produces distinct temporal characteristics. When triggered at $t=2s$, the attack induces immediate local voltage disturbance propagating through the network according to a voltage decay profile governed by network impedance characteristics and load response dynamics. After the initial voltage fluctuations, the main bus 632 voltage begins to collapse after $t=60s$ due to the grid's lack of active \& reactive power to maintain voltage stability, ultimately causing a blackout. The frequency also drops to zero as the system becomes unstable.

III) Asymmetric output generation from compromised DC-AC stage operation, modeled as progressive loss of excitation combining voltage irregularities and synchronization issues, capturing both the voltage instability and reactive power implications of asymmetric inverter operation.

The DC-AC stage attack (Fig.~\ref{fig:ETAP_Voltage_Impact_Fault}(C)) demonstrates similar instability patterns, with the main bus experiencing synchronization issues earlier than in the DC-DC attack scenario. This rapid progression results from the asymmetric waveform distortion's immediate impact on grid synchronization, creating more severe reactive power imbalances.

\begin{figure*}[t!]
    \centering
    \includegraphics[trim={0px 10px 0px 20px},width=0.86\linewidth]{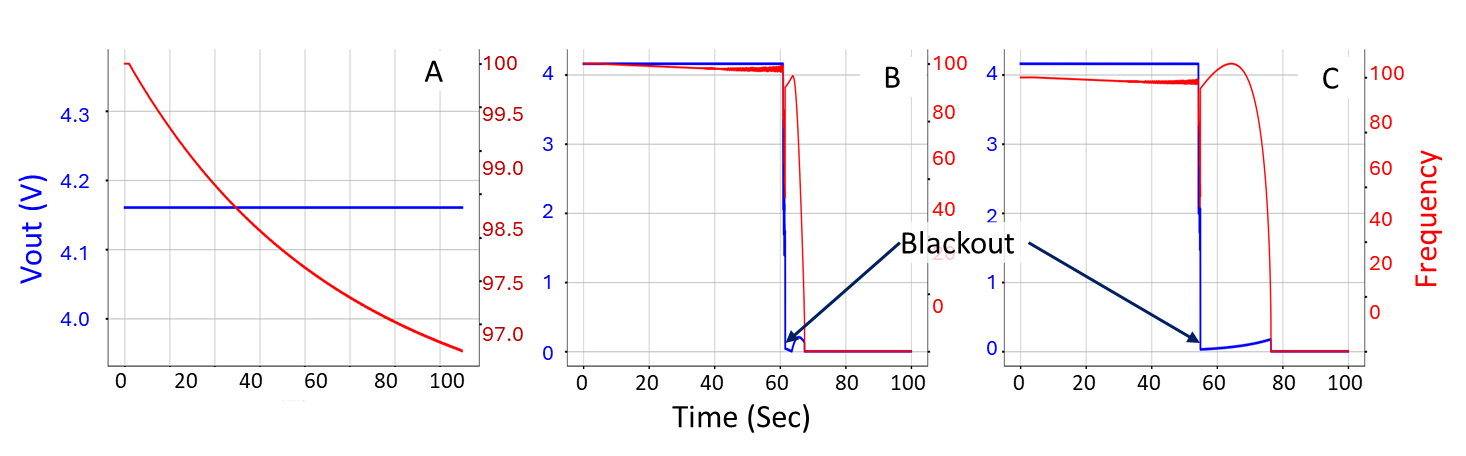}
    \caption{ETAP simulation results showing grid response to HT-induced faults: (A) Inverter Disconnection Fault, (B) Voltage Instability Propagation Fault, and (C) Inverter-Induced Synchronization Fault.}
\label{fig:ETAP_Voltage_Impact_Fault}
    % \vspace{-5pt}
\end{figure*}

\subsubsection{Cascade Progression and System Response}
The failure mechanism follows a characteristic sequence: the compromised 100 kW generator creates an immediate power deficit that other generators in the system must compensate. During peak demand, when generators operate at capacity limits, they cannot provide adequate compensation. 
The resulting deficit in reactive power supply initiates a cascade through the Q-V (reactive power-voltage) relationship. Initial voltage drops impair the grid's ability to transfer reactive power, which in turn leads to further voltage degradation. This creates a self-reinforcing cycle: reduced voltage levels progressively diminish reactive power transmission capability, accelerating the system's descent into voltage collapse. Our ETAP simulations confirm this behavior across all attack vectors.

\section{Discussion}
\label{sec:discussion}
\subsection{Attack Generalization}
While we demonstrated ERM using temperature sensors, the attack principle extends to other sensing systems in power electronics. The key vulnerability lies in the common signal conditioning architecture - specifically, output filtering capacitors that enable our rate-detection approach through $dV/dt$ monitoring. 
For instance, hall-effect sensors used for current measurement in solar inverters share similar vulnerabilities. Barua et al.~\cite{barua_hall_2020} previously demonstrated these sensors' vulnerability to external magnetic field attacks, leading to subsequent defense proposals involving redundant hall sensors~\cite{halc_barua}. 
However, such redundancy-based defenses remain vulnerable to ERM attacks due to a fundamental difference in attack mechanism as discussed in Sec.~\ref{sec:threat_model}. Our approach can be adapted to any sensor with similar signal conditioning, requiring only adjustments to match specific sensor characteristics.

\subsection{Detection Evasion and Mitigation Implications} 
\label{subsec:evasion_detection_mitigation}

A critical aspect of the ERM Trojan is its ability to evade a wide array of existing hardware Trojan detection techniques. As detailed in Section~\ref{sec:integration_minimization_detection}, the Trojan's extremely low power consumption (sub-nanoWatt), minimal timing delay (around 114\,ps), small physical footprint ($14~\mu\text{m}^2$), and novel rate-based triggering mechanism collectively render it highly resistant to detection by side-channel analysis, optical inspection, standard functional testing, and run-time monitoring. This challenges the assumption that Trojans must introduce significant, detectable overheads or rely on easily replicable trigger conditions. The fact that ERM operates by monitoring the \textit{rate of change} of environmental parameters, rather than their absolute values exceeding a threshold, means it can remain dormant even during comprehensive testing across different operational conditions if those tests do not specifically vary conditions at the \textit{rate} the Trojan is sensitive to. This highlights a potential blind spot in current hardware security validation processes that primarily focus on static or slowly varying conditions.

\subsection{Limitations}
Our hardware validation was limited by the catastrophic nature of the attack's impact. When testing the DC-DC converter attack scenario, the immediate failure of the photovoltaic full-bridge driver chip and subsequent component damage prevented us from conducting extensive hardware-in-the-loop experiment of alternative attack scenarios. While this demonstrated the attack's potency, it limited our ability to fully validate all theoretical attack vectors in hardware.

Furthermore, while we build upon established temperature sensor vulnerabilities demonstrated in prior work~\cite{tu2019trick}, our analysis assumes the attacker can achieve the same level of precise EMI control. Additionally, our grid-level analysis, though comprehensive in simulation, would benefit from validation in larger-scale power system testbeds to capture complex interactions in real power grids with diverse generation sources and protection schemes.

\color{black}

\section{Conclusion}
\label{sec:conclusion}
This paper introduces ERM, a novel HT triggering mechanism that monitors rate of change in environmental sensor readings rather than specific conditions or patterns. This approach exploits a fundamental security gap by operating within normal parameter ranges while responding to specific rates of change, bypassing current detection methods. Our comprehensive evaluation, from theoretical analysis to hardware validation and grid-scale simulation, demonstrates how a single compromised 100kW inverter can trigger cascading failures through reactive power-voltage interactions, ultimately causing widespread outages. The attack's minimal circuit footprint (14$\mu$m$^2$), use of existing sensor infrastructure, and ability to circumvent redundancy-based defenses present significant challenges for hardware security as power systems increasingly depend on inverter-based resources.

\newpage

\bibliographystyle{IEEEtran}
\bibliography{IEEEabrv,bibfile}
\appendices
\section{Evaluation}
\begin{figure}[h!]
    \centering
    \includegraphics[width=\linewidth]{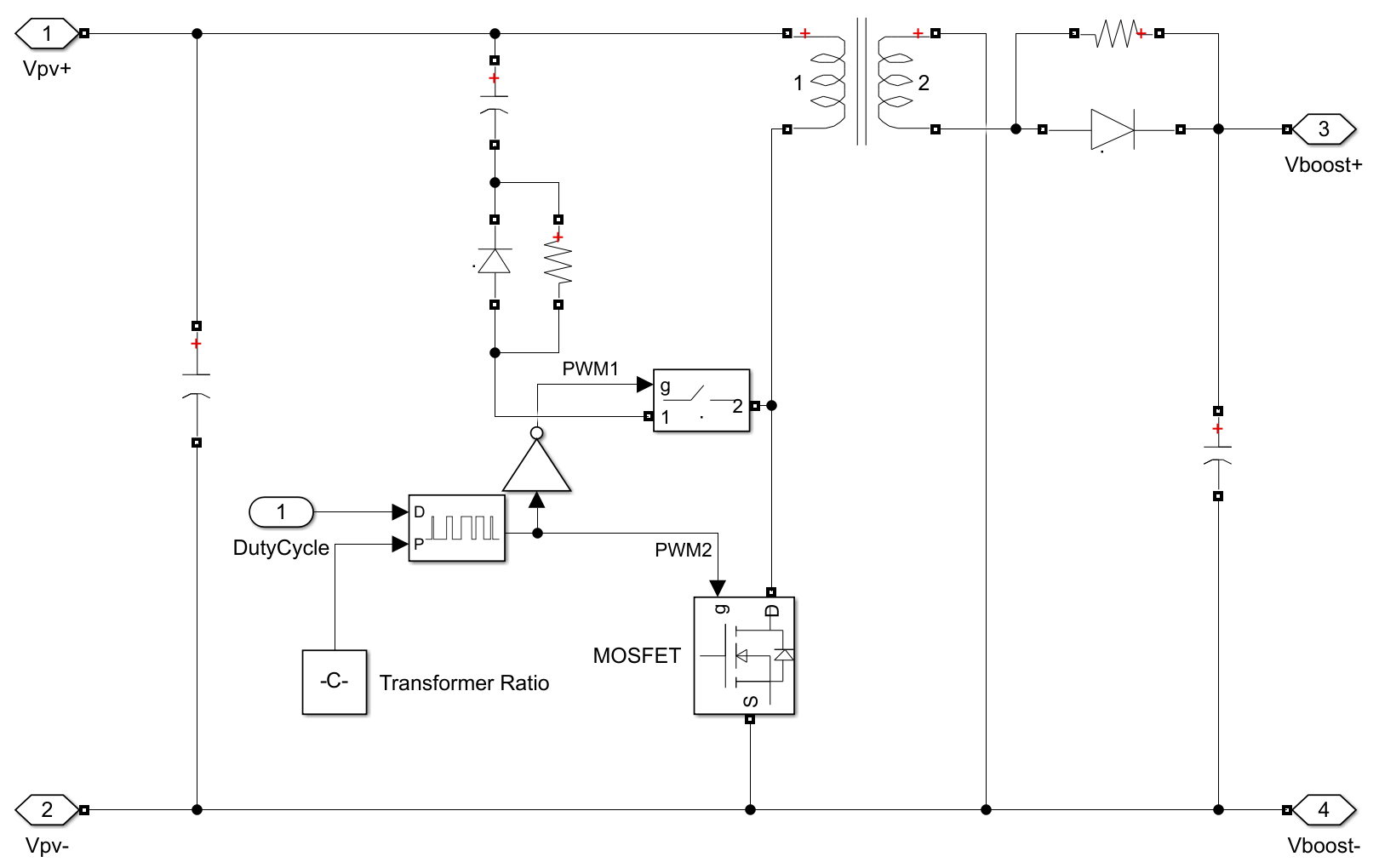}
    \caption{Simulink DC-DC Converter Model}
    \label{fig:Simulink_DC-DC_Converter_Model}
\end{figure}
\begin{figure}[h!]
    \centering
    \includegraphics[width=0.8\linewidth]{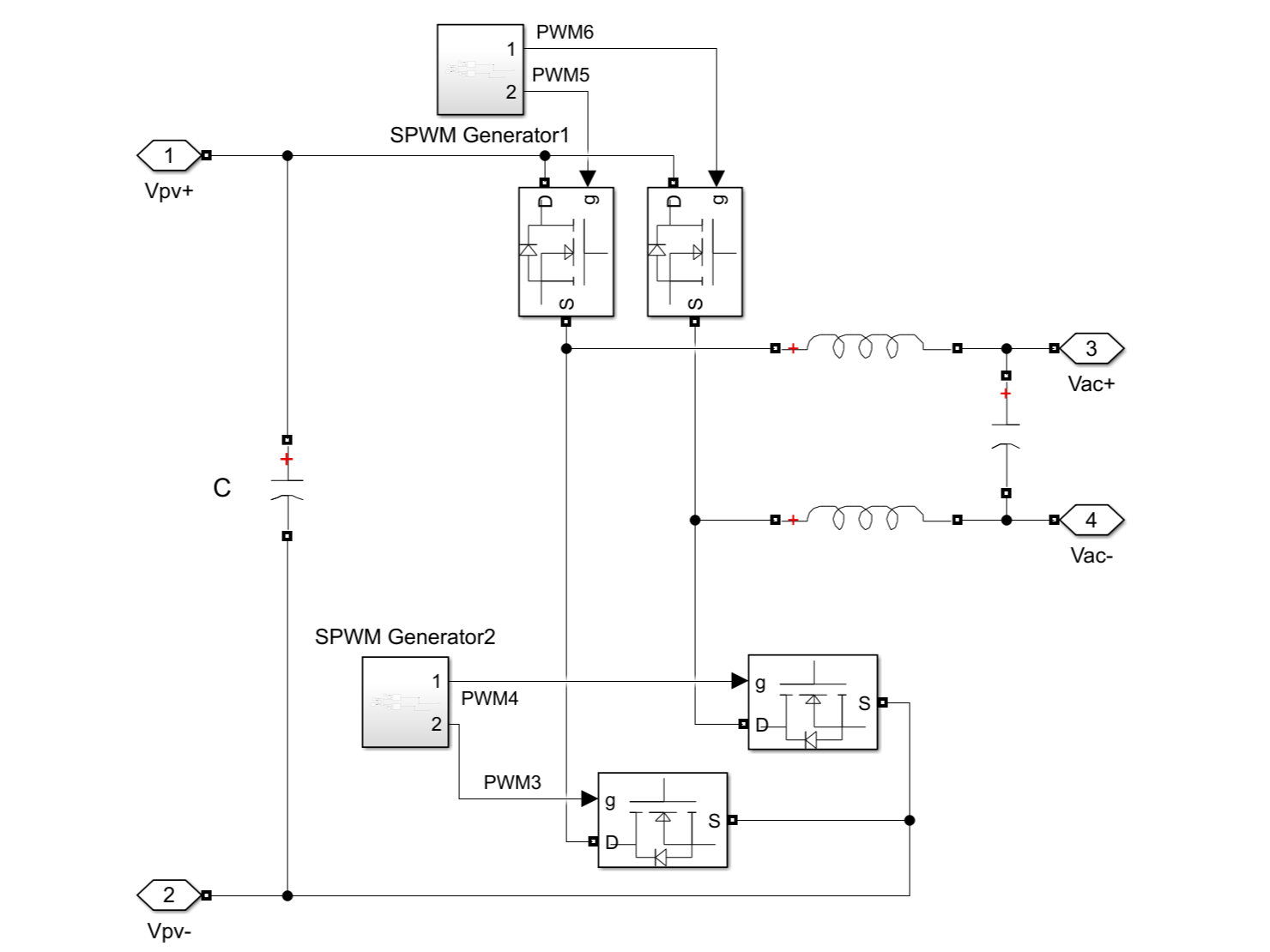}
    \caption{Simulink DC-AC Inverter Model}
    \label{fig:Simulink_DC-AC_Inverter_Model}
\end{figure}
\subsection{Complete Solar Inverter Simulation}
\subsubsection{Simulink DC-DC Converter Model} 
\label{appx:simulink_dcdc}
Fig.~\ref{fig:Simulink_DC-DC_Converter_Model} depicts the DC-DC boost circuit in our Simulink model.

\subsubsection{Simulink DC-AC Converter Model}
\label{appx:simulink_dcac}
Fig.~\ref{fig:Simulink_DC-AC_Inverter_Model} depicts the DC-DC inverter circuit in our Simulink model.

\end{document}